\documentclass{llncs}
\usepackage[utf8]{inputenc}
\usepackage[T1]{fontenc}
\usepackage[scaled=0.8]{beramono}
\usepackage{amsmath}
\usepackage{amssymb}
\usepackage{stmaryrd}
\usepackage{mathtools}
\usepackage{proof}
\usepackage{todonotes}
\usepackage{subcaption}
\usepackage{wrapfig}
\usepackage{sidecap}
\usepackage{xspace}
\usepackage{booktabs}
\usepackage{multirow}

\usepackage{tabto}

\usepackage{listings}
\newcommand{\Korn}{\textsc{Korn}\xspace}

\usepackage{newunicodechar}
\usepackage{tikz}
\usetikzlibrary{snakes}

\title{A Complete Approach to Loop Verification \\ with Invariants and Summaries
    \thanks{This article extends a conference version at VMCAI'22 with an evaluation (\cref{sec:evaluation}).}}
\author{Gidon Ernst}
\institute{LMU Munich, \email{gidon.ernst@lmu.de}}

\usepackage[numbers,sort&compress]{natbib}
\usepackage{comment}
\usepackage{hyperref}
\hypersetup{hidelinks,
    colorlinks=true,
    allcolors=blue,
    pdfstartview=Fit,
    pdftitle=Loop Verification with Contracts,
    breaklinks=true}

\usepackage[capitalise]{cleveref}

\Crefname{equation}{Eq.}{Eqs.}
\Crefname{figure}{Fig.}{Figs.}
\Crefname{tabular}{Table}{Tabs.}
\Crefname{example}{Ex.}{Exs.}
\Crefname{section}{Sec.}{Sects.}
\Crefname{corollary}{Cor.}{Cors.}
\Crefname{proposition}{Prop.}{Props.}
\Crefname{theorem}{Thm.}{Thms.}
\Crefname{lemma}{Lem.}{Lems.}
\Crefname{algorithm}{Alg.}{Algs.}
\Crefname{definition}{Def.}{Defs.}

\renewcommand{\paragraph}[1]{\smallskip\noindent\textbf{#1~}}

\usepackage{newunicodechar}

\newcommand{\cupdot}{\mathbin{\mathaccent\cdot\cup}}
\newcommand{\usym}[2]{\newunicodechar{#1}{\ensuremath{#2}}}

\mathchardef\mhyphen="2D

\def\comp  {\mathbin{\raise 0.6ex\hbox{\oalign{\hfil$\scriptscriptstyle
    \mathrm{o}$\hfil\cr\hfil$\scriptscriptstyle\mathrm{9}$\hfil}}}}

\usym{○}{\circ}
\usym{□}{\Box}

\usym{ŝ}{\hat s}
\usym{Ŝ}{\hat S}

\usym{⁺}{\ensuremath{^+}}
\usym{⁻}{\ensuremath{^-}}
\usym{ⁱ}{\ensuremath{^i}}
\usym{ⁿ}{\ensuremath{^n}}
\usym{⁰}{\ensuremath{^0}}
\usym{¹}{\ensuremath{^2}}
\usym{²}{\ensuremath{^2}}
\usym{³}{\ensuremath{^3}}
\usym{⁴}{\ensuremath{^4}}
\usym{⁵}{\ensuremath{^5}}
\usym{⁶}{\ensuremath{^6}}
\usym{⁷}{\ensuremath{^7}}
\usym{⁸}{\ensuremath{^8}}
\usym{⁹}{\ensuremath{^9}}

\usym{₊}{\ensuremath{_+}}
\usym{₋}{\ensuremath{_-}}
\usym{ᵢ}{\ensuremath{_i}}
\usym{ₖ}{\ensuremath{_{i+1}}}
\usym{ₙ}{\ensuremath{_n}}
\usym{₀}{\ensuremath{_0}}
\usym{₁}{\ensuremath{_1}}
\usym{₂}{\ensuremath{_2}}
\usym{₃}{\ensuremath{_3}}
\usym{₄}{\ensuremath{_4}}
\usym{₅}{\ensuremath{_5}}
\usym{₆}{\ensuremath{_6}}
\usym{₇}{\ensuremath{_7}}
\usym{₈}{\ensuremath{_8}}
\usym{₉}{\ensuremath{_9}}

\usym{∥}{|\!|}
\usym{⊥}{\bot}
\usym{∩}{\cap}
\usym{·}{\cdot}
\usym{…}{\ldots\xspace} 
\usym{⋯}{\cdots\xspace}
\usym{∘}{\circ}
\usym{∷}{\colon}
\usym{≔}{\coloneqq}
\usym{≡}{\equiv}
\usym{≈}{\approx}
\usym{∪}{\cup}
\usym{⋄}{\Diamond}
\usym{↓}{\downarrow}
\usym{∅}{\varnothing}
\usym{∃}{\exists\ }
\usym{∀}{\forall\ }
\usym{≥}{\ge}
\usym{⊓}{\glb}
\usym{⟫}{\rangle\!\rangle}
\usym{∈}{\in}
\usym{∞}{\infty}
\usym{∫}{\int}
\usym{⊔}{\join}
\usym{⟨}{\langle}
\usym{⦉}{\langle\!|}
\usym{⦊}{|\!\rangle}
\usym{⌈}{\lceil}
\usym{≤}{\le}
\usym{←}{\leftarrow}
\usym{⇐}{\Leftarrow}
\usym{↔}{\leftrightarrow}
\usym{⇔}{\Leftrightarrow}
\usym{⌊}{\lfloor}
\usym{↯}{\lightning}
\usym{⟦}{\llbracket}
\usym{⌜}{\lq}
\usym{⟪}{\langle\!\langle}
\usym{↦}{\mapsto}
\usym{⊧}{\models}
\usym{⊭}{{\not\models}}
\usym{¬}{\lnot}
\usym{≠}{\neq}
\usym{∉}{\notin}
\usym{⊙}{\odot}
\usym{⊖}{\ominus}
\usym{⊕}{\oplus}
\usym{⊘}{\oslash}
\usym{⊗}{\otimes}
\usym{⇸}{\pto}
\usym{⟩}{\rangle}
\usym{⌉}{\rceil}
\usym{⌋}{\rfloor}
\usym{→}{\rightarrow}
\usym{⇒}{\implies}
\usym{⌝}{\rq}
\usym{⟧}{\rrbracket}
\usym{◪}{\smallanysquare}
\usym{■}{\smallblacksquare}
\usym{□}{\smallsquare} 
\usym{⊑}{\sqsubseteq}
\usym{⊏}{\sqsubset}
\usym{⊒}{\sqsupseteq}
\usym{⊐}{\sqsupset}
\usym{⊂}{\subset}
\usym{⊆}{\subseteq}
\usym{⊃}{\supset}
\usym{⊇}{\supseteq}
\usym{×}{\times}
\usym{⊤}{\top}
\usym{△}{\triangle}
\usym{◁}{\triangleleft}
\usym{▷}{\triangleright}
\usym{↑}{\kern 1pt \raisebox{3pt}{\scalebox{0.9}{$\uparrow$}}}
\usym{⊦}{\vdash}
\usym{∨}{\vee}
\usym{∧}{\wedge}
\usym{α}{\alpha}
\usym{β}{\beta}
\usym{Γ}{\Gamma}
\usym{γ}{\gamma}
\usym{Δ}{\Delta}
\usym{δ}{\delta}
\usym{ε}{\epsilon}
\usym{ζ}{\zeta}
\usym{η}{\eta}
\usym{θ}{\theta}
\usym{ι}{\iota}
\usym{κ}{\kappa}
\usym{Λ}{\Lambda\ }
\usym{λ}{\lambda\ }
\usym{μ}{\mu}
\usym{ν}{\nu}
\usym{Ξ}{\Xi}
\usym{ξ}{\xi}
\usym{Π}{\Pi}
\usym{π}{\pi}
\usym{ρ}{\rho}
\usym{Σ}{\Sigma}
\usym{σ}{\sigma}
\usym{τ}{\tau}
\usym{Φ}{\Phi}
\usym{φ}{\phi}
\usym{χ}{\chi}
\usym{Ψ}{\Psi}
\usym{ψ}{\psi}
\usym{Ω}{\Omega}
\usym{ω}{\omega}
\usym{ℕ}{\mathbb{N}}
\usym{𝔹}{\mathbb{B}}
\usym{ℙ}{\mathbb{P}}
\usym{¦}{\pipe}
\usym{≃}{\simeq}
\usym{⊎}{\uplus}
\usym{⊍}{\cupdot}
\usym{★}{\mathrel{\star}}
\usym{⟅}{\lbag}
\usym{⟆}{\rbag}
\usym{ῠ}{\breve\upsilon}
\usym{ῐ}{\breve\iota}
\usym{ς}{\varsigma}
\usym{⋈}{\bowtie}

\definecolor{darkgreen}{rgb}{0,0.6,0}
\renewcommand{\qed}{\hfill\ensuremath{\square}}
\newcommand{\blue}[1]{{\color{blue}#1}}

\newcommand{\red}[1]{{\color{red}#1}}

\newcommand{\ol}[1]{\overline{#1}}

\newcommand{\err}{\lightning}
\newcommand{\brk}[1]{{#1}_\downarrow}

\newcommand{\pipe}{\mathrel{\makebox[\widthof{$\Coloneqq$}]{$|$}}} 

\newcommand{\xs}{\ol x}
\newcommand{\ys}{\ol y}

\renewcommand{\phi}{\varphi}

\newcommand{\sorted}{\mathit{sorted}}
\renewcommand{\max}{\mathit{max}}

\newcommand{\False}{\mathit{false}}

\newcommand{\code}[1]{\texttt{\upshape #1}}
\newcommand{\hoare}[3]{\{\,#1\,\}\,#2\,\{\,#3\,\}}

\newcommand{\safe}[2]{\{\,#1\,\}\,#2~\textit{is safe}}
\newcommand{\correct}[3]{\{\,#1\,\}\,#2\,\{\,#3\,\}~\textit{is correct}}

\newcommand{\old}{\code{old}}

\newcommand{\cmd}[1]{\code{#1}}

\newcommand{\SPEC}[3]{#1\colon[#2,#3]}

\newcommand{\DO}{\cmd{do}}

\newcommand{\SKIP}{\cmd{skip}}
\newcommand{\WHILE}{\cmd{while}}

\newcommand{\BREAK}{\cmd{break}}

\newcommand{\GOTO}{\cmd{goto}}

\begin{document}
\maketitle

\begin{abstract}
\emph{Invariants} are the predominant approach to verify the correctness of loops.
As an alternative, \emph{loop contracts}, which make explicit the premise and conclusion of the underlying induction proof,
can sometimes capture correctness conditions more naturally.
But despite this advantage, the second approach receives little attention overall,
and the goal of this paper is to lift it out of its niche.
We give the first comprehensive exposition of the theory of loop contracts,
including a characterization of its completeness.
We show concrete examples on standard algorithms that showcase their relative merits.
Moreover, we demonstrate a novel constructive translation between the two approaches,
which decouples the chosen specification approach from the verification backend.

\end{abstract}

\section{Introduction}
\label{sec:introduction}

Loop \emph{invariants}~\cite{hoare1969axiomatic} are the standard approach to verify programs with loops.
The technique is practically successful for both specifying and verifying loops in automated tools.
The corresponding proof obligations propagate invariants forwards over a single arbitrary iteration, and soundness is justified by the induction principle of the least fixpoint of the loop.

The alternative approach is to specify loops in terms of a \emph{contract} consisting of a precondition and a relational postconditon (called ``summary'' here), as advocated e.g. by
Hehner~\cite{hehner2005specified,hehner1999refinement}.
Contracts have two important features:
1) they tend to resemble the overall program specification more closely when compared to their plain invariant counterparts,
and 2) they can dually express proof arguments that propagate backwards from the result.
Essentially, loops are treated analogously to tail-recursive procedures,
but without the need for an explicit syntactic translation.
The benefits of such flexible proof schemas for loops are widely acknowledged, e.g.~\cite{bohorquez2010elementary,rosu2009circular,chargueraud2010characteristic},
notably in Separation Logic, where tracking the frame is problematic with just invariants~\cite{tuerk2010local,jacobs2015solving,ernst:sttt2015,brotherston2005cyclic}.

Surprisingly, while contracts have been described in the literature and implemented in tools such as VeriFast~\cite{philippaerts2014software},
the theoretical connection between invariants, loop pre- and postconditions,
as well as completeness of the contract approach appear to be unresolved.
Moreover, examples tend to be given in the context of Separation Logic
but not for standard verification problems.

\paragraph{Contribution and Outline:} 
In this paper, we provide a deep investigation of loop contacts in comparison to invariants,
from a theoretical and from an empirical point of view, leading to the following technical results:
\begin{itemize}
\item
    We formulate contract-based verification (\cref{sec:approach})
    to clearly exhibit the coincidence of invariants and loop preconditions,
    and the dual nature of invariants and loop summaries.
    Thereby we generalize Hoare's approach~\cite{hoare1969axiomatic};
    as well as Hehner and Gravel's technique for \code{for}-loops~\cite{hehner1999refinement}
    to all \code{while}-loops.
\item
    Just as variants capture the delta between partial and total correctness,
    loop preconditions correspond to absence of runtime errors in the loop body,
    leading to yet a weaker correctness criterion for which
    loop summaries alone give a \emph{complete} verification method (\cref{thm:cc}).
\item
    We provide \emph{constructive translations} between plain invariants and loop contracts (\cref{prop:lift,prop:lower} in \cref{sec:bridge}),
    which explains their parity and moreover provides key guidelines for building and integrating tools.
\item
    We reify Tuerk's approach~\cite{tuerk2010local} as a syntactic proof
    rule (\cref{sec:hoare}) that that lends itself directly for implementation in typical
    verification tools,
    by leveraging specification statements~\cite{morgan1988specification}.
\end{itemize}
The key take-away is that contracts offer a particular
and useful way to think about the correctness of loops,
that is \emph{conceptually} different from invariants,
but at the same time,
the \emph{technical} requirements for supporting this approach
turn out to be superficial, and tool support is straight-forward.

As a consequence, we are at liberty to choose the approach that fits a particular problem most.
But what does that mean in practice?
What are the advantages and disadvantages of contracts in comparison to invariants?
\begin{itemize}
\item
    We specify the correctness of a number of well-known algorithms with contracts (\cref{sec:examples}),
    characterize when and why loop summaries may carry the bulk of the proof,
    and also give insights into their limitations.
\end{itemize}
We show that loop contracts may resemble
the respective correctness requirements more closely and require minor generalizations only,
when compared to their invariant counterparts.
Loop summaries are suitable for those properties,
which naturally propagate backwards and are thus misaligned with the forward computation of a loop.
On the other hand, they tend to require additional frame conditions
to preserve modifications of data structures across iterations.

Finally, we are interested in proof automation:
\begin{itemize}
\item
    We evaluate whether state-of-the-art Horn clause solvers~%
        \cite{grebenshchikov2012synthesizing,bjorner2015horn,hojjat2018eldarica}
    are effective not only in instatiating invariants but also loop contracts (\cref{sec:evaluation})
    in a side-by-side comparison on SV-COMP benchmarks~\cite{beyer2020advances}.
\end{itemize}
Primary goal of this evaluation test whether the completeness result carries practical significance.
The results, while promising, may reflect a bias towards invariants in existing tools:
Performance of Eldarica~2.0.4 and Z3~4.8.9 with contracts is within 90\% resp. 70\% of that with invariants.

\paragraph{Proofs:} A mechanization in Isabelle/HOL~\cite{nipkow2002isabelle}
of the theory presented in \cref{sec:approach,sec:bridge}
is available at \url{https://zenodo.org/record/5509953}.

\section{Motivation and Overview}
\label{sec:motivation}

In this section we exemplify proofs using invariants and proofs using loop contracts.
The running example is Challenge~1 from VerifyThis 2011~\cite{verifythis2011}:
finding the maximum in an array by elimination, as shown in \cref{fig:max}.
The program maintains a subrange $\code{a}[\code{l}..\code{r}+1]$ wrt. two indices $\code{l}$ and $\code{r}$
of candidates for the maximum in in array~$\code{a}$ of length~$\code{n}$.
In each iteration, the smaller of the two candidates is eliminated
from the subrange, either by incrementing~$\code{l}$ or by decrementing~$\code{r}$.
Correctness of the algorithm depends on the fact that the maximum remains in that range.
The specification, annotated at the top left expresses that the return value, denoted $\mathbf{res}$,
is equal to the result given by logic function $\max$, where 
$\code{a}[0..\code{n}]$ denotes the non-empty sequence of array elements at indices $0,\ldots,\code{n}-1$.

\begin{figure}[t]
\centering
    \begin{minipage}{0.43\textwidth}
\begin{lstlisting}
int max(int a[], int n)
  requires $0 < \code{n}$
  ensures  $\code{a}[\mathbf{res}] = \max(\code{a}[0..\code{n}])$
{
  int l = 0;
  int r = n-1;
  while(l != r)
    if(a[l] <= a[r]) l = l+1;
    else             r = r-1;
  return l;
}
\end{lstlisting}
    \end{minipage}
\quad
    \begin{minipage}{0.52\textwidth}
    \textbf{invariant} (Filliâtre \& Marché) \\[2pt]
    $\bullet$ $0 ≤ \code{l} ≤ \code{r} < \code{n}$ \\
    $\bullet$ $∀ k.\ 0 ≤ k < \code{l} ∨ \code{r} < k < \code{n}$ \\
              \hspace*{0.8cm} ${} \implies \code{a}[k] ≤ \max(\code{a}[\code{l}], \code{a}[\code{r}])$


    \medskip

    \textbf{invariant} (Ernst, Schellhorn, Tofan) \\[2pt]
    $∃ k.\ 0 ≤ \code{l} ≤ k ≤ \code{r} < \code{n} ∧ \code{a}[k] = \max(\code{a}[0..\code{n}])$

    \medskip

    \textbf{contract} \\
    precondition:  \tabto{2.5cm} $0 ≤ \code{l} ≤ \code{r} < \code{n}$ \\
    summary: \tabto{2.5cm} $\code{a}[\code{l}'] = \max(\code{a}[\code{l}..\code{r}+1])$ \\
    (primed variables refer to loop exit)
    \end{minipage}

    \caption{Finding the maximum element in an array by elimination.
             Algorithm (left) and different correctness arguments (right),
             which are all sufficient alone.}
    \label{fig:max}
\end{figure}

\paragraph{Specification.}
At the top-right of \cref{fig:max}, two example \emph{invariants} are shown.
The one used by the Why~3 team in \cite{verifythis2011}
imposes an ordering on the index variables (first $\bullet$)
and expresses that both the left part and the right part
of the array contains elements that have been rightfully excluded,
i.e., one of the two boundary values is greater than all of these
(second $\bullet$).
The invariant discovered by the KIV team expresses that
there remains some index~$k$ within the range that is maximum of the whole array,
implying that this range remains non-empty for the maximum to be well-defined.
In both cases, the program's postcondition follows from the invariant
when the loop guard becomes false, i.e, when $\code{l} = \code{r}$.

We can alternatively specify the loop using a \emph{contract},
which consists of a \emph{loop precondition}, later called a ``safe'' invariant (cf. \cref{def:inv}), whose role is to guarantee that the loop executes without error,
and a \emph{summary}, which establishes functional correctness.
The latter is a relation between current unprimed values and primed final values
of the program variables.
From whichever intermediate indices $\code{l}$ and $\code{r}$ we jump into the execution of the loop,
the element at the final index $\code{l}'$ will be maximal for that subrange.
The specification of the program is implied for $\code{l}' = \mathbf{res}$
wrt. initial values $\code{l} = 0$ and $\code{r} = \code{n}-1$.
The loop precondition is about the ordering of indices
to keep track that the range remains nonempty.

Note that first, the contract reflects the intuition behind the algorithm more naturally:
It computes a final index $\code{l}'$ that points to the maximum.
Moreover, this summary occurs almost verbatim in the annotation of procedure $\code{max}$, albeit with the fixed bounds~$0$ and $\code{n}$ that have to be generalized (such generalization is, of course, unavoidable).
In \cref{sec:bridge} we show a third possibility to state an invariant,
motivated by and constructed from this summary.

\begin{figure}[t]
    \centering
    \begin{tikzpicture}[node distance=2cm]
\node (s0) {$s_0$};
\node (si) [right=2.5cm of s0] {$s_{i}$};
\node (sj) [right=1.5cm of si] {$s_{i+1}$};
\node (sn) [right=2.8cm of sj] {$s_{n}$};

\draw[->,thick,dashed] (s0) -- (si);
\draw[->,thick] (si) -- node[above] {body} (sj);
\draw[->,thick,dashed] (sj) -- (sn);

\node (P0) [above=0.35cm of s0] {$\blue{P(s_0)} \implies \red{I(s_0)}$};
\node (Ii) [above=0.35cm of si] {$\blue{I(s_i) \land t(s_i)}$};
\node (Ij) [above=0.35cm of sj] {$\implies \red{I(s_{i+1})}$};
\node (Qn) [above=0.35cm of sn] {$\blue{I(s_n) \land \lnot t(s_n)} \implies \red{Q(s_n)}$};

\draw[thick,dotted] (P0) -- (s0);
\draw[thick,dotted] (Ii) -- (si);
\draw[thick,dotted] (Ij) -- (sj);
\draw[thick,dotted] (Qn) -- (sn);

    \end{tikzpicture}
    \caption{Forward propagation of invariant~$I$.
             Blue: assumptions, red: to prove.}
    \label{fig:inv}
\end{figure}
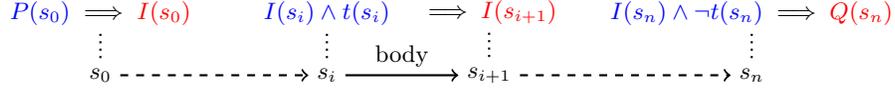

\paragraph{Proofs.}
Schematically, we can describe the execution of the loop,
using logical variables~$lᵢ$ and~$rᵢ$
indexed by the $i$-th iteration, for $i = 0,\ldots,n$,
where $l₀ = 0$, $r₀ = \code{n}-1$, and $lₙ = rₙ$.
We denote by $I(lᵢ,rᵢ)$ the instantiation of (either of the) invariants
with $\code{l},\code{r} \mapsto lᵢ,rᵢ$.
Similarly, $R(lᵢ,rᵢ,lₙ,rₙ)$ is the instantiation of the summary,
with $\code{l},\code{r} \mapsto lᵢ,rᵢ$
and $\code{l}',\code{r}' \mapsto lₙ,rₙ$ (note that $\code{r}'$ does not occur).
Hence, $I$ describes all states encountered loop head,
including when the loop is entered first and right when the loop exits.
$R$~describes the relation between these states at loop head
and the final states at loop exit.

The proof that an invariant~$I$ is correct considers the usual three conditions,
visualized in \cref{fig:inv}.
$I(l₀,r₀)$ holds initially, $I(lᵢ,rᵢ) ⇒ I(lₖ,rₖ)$ propagates forwards through iterations for each $i$,
and $I(lₙ,rₙ) ⇒ Q(lₙ,rₙ)$ finally establishes the postcondition~$Q$ of \code{max}.
With some knowledge about function $\max$,
it is easy to check that both invariants shown in \cref{fig:max} satisfy these conditions.

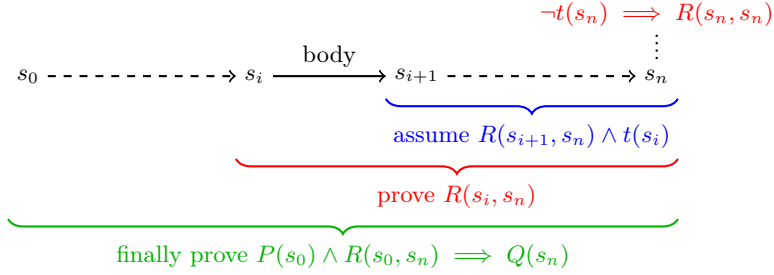
\begin{figure}[t]
    \centering
    \begin{tikzpicture}[node distance=2cm]
\node (s0) {$s_0$};
\node (si) [right=2.5cm of s0] {$s_{i}$};
\node (sj) [right=1.5cm of si] {$s_{i+1}$};
\node (sn) [right=2.5cm of sj] {$s_{n}$};

\draw[->,thick,dashed] (s0) -- (si);
\draw[->,thick] (si) -- node[above] {body} (sj);
\draw[->,thick,dashed] (sj) -- (sn);

\node (Qn) [above=0.35cm of sn] {$\red{\lnot t(s_n) \implies R(s_n,s_n)}$};

\draw[thick,dotted] (Qn) -- (sn);

\draw[thick,decorate,decoration={brace,raise=0.3cm,amplitude=2mm,mirror},blue]
    (sj.west) -- (sn.east)
    node [pos=0.5,yshift=-0.8cm] {assume $R(s_{i+1},s_n) \land t(s_i)$}; ;

\draw[thick,decorate,decoration={brace,raise=1.1cm,amplitude=2mm,mirror},red]
    (si.west) -- (sn.east)
    node [pos=0.5,yshift=-1.6cm] {prove $R(s_i,s_n)$}; ;

\draw[thick,decorate,decoration={brace,raise=1.9cm,amplitude=2mm,mirror},green!70!black]
    (s0.west) -- (sn.east)
    node [pos=0.5,yshift=-2.4cm] {finally prove $P(s_0) \land R(s_0,s_n) \implies Q(s_n)$}; ;
    \end{tikzpicture}
    \caption{Backward propagation conditions for a loop postcondition~$R$
             with respect to an iteration of the loop.
             Whe overall conclusion is marked green.}
    \label{fig:sm}
\end{figure}

Dually, the proof that $R$ is indeed a correct summary also considers three conditions,
as visualized in \cref{fig:sm}:
We need to check that $R(lₙ,rₙ)$ holds finally, i.e., the poscondition adequately summarizes
the computation of a loop that terminates immediately,
$R(lₖ,rₖ,lₙ,rₙ) ⇒ R(lᵢ,rᵢ,lₙ,rₙ)$ for all~$i$ ensures that an additional leading iteration is covered, too.
As a result, the entire computation of the loop is summarized by~$R$,
from which we need to establish the postcondition of~\code{max} by
$R(l₀,r₀,lₙ,rₙ) ⇒ Q(lₙ,rₙ)$.

We briefly sketch the critical step of backwards propagation for the~$R$ shown in \cref{fig:max}, in the case that \code{a[l] <= a[r]} evaluates to true in the $i$-th iteration.
Specifically, given that $lₖ = lᵢ + 1$ and $rₖ = rᵢ$, from
\begin{align}
R(lₖ,rₖ,lₙ,rₙ)
    \quad &≡ \quad \code{a}[lₙ] = \max(\code{a}[lₖ..rₖ+1])
    \label{eq:R'}
\end{align}
we propagate $R$ from state $i+1$ back to state~$i$, where
\begin{align}
R(lᵢ,rᵢ,lₙ,rₙ)
    \quad &≡ \quad \code{a}[lₙ] = \max(\code{a}[lᵢ..rᵢ+1]).
    \label{eq:R}
\end{align}
Using equality
$\max(\code{a}[lᵢ..rᵢ+1]) = \max(\code{a}[lᵢ], \max(\code{a}[lᵢ+1..rₖ+1]))$
and substituting variables,
it remains to be proven that $\code{a}[lₙ] = \max(\code{a}[lᵢ], \code{a}[lₙ])$,
i.e., $\code{a}[lᵢ] ≤ \code{a}[lₙ]$,
which follows by transitivity from the \code{if} condition $\code{a}[lᵢ] ≤ \code{a}[rᵢ]$
and $\code{a}[rᵢ] ≤ \code{a}[lₙ]$ as a consequence of \eqref{eq:R'}.
We remark that these reasoning steps are easy for automatic provers
when provided with the obvious properties of $\max$.

\section{Preliminaries}
\label{sec:preliminaries}

We consider imperative commands~$C$, defined over a semantic domain of states~$S$,
as relations $C ⊆ S × Ŝ$ with
$Ŝ = S ⊎ \{\err\} ⊎ \{ \brk s \mid s ∈ S \}$,
where~$\err$ signifies a runtime error (e.g. failed assertion, division by zero, out of bounds array access),
and $\brk s$ signifies early exit of a loop via a $\BREAK$ command in state~$s$.
By notational convention~$ŝ ∈ Ŝ$, whereas~$s ∈ S$ strictly.
Nontermination, which is orthogonal to this paper, is reflected by the absence of successor states as usual.
We use suggestive naming, e.g. state~$s_0 ∈ S$ is used for the initial state of some execution, $s,s',s''$ are intermediate states, and $sₙ$ typically refers to a final state of the loop.

\begin{definition}[Validity of Hoare-Triples]
    \label{def:hoare}
A command~$C ⊆ S × Ŝ$ is partially correct wrt. a precondition~$P ⊆ S$ and a postcondition~$Q ⊆ S$, if the Hoare triple $\hoare{P}{C}{Q}$ is valid, written $⊧ \hoare{P}{C}{Q}$, and defined as usual
\[ ⊧ \hoare{P}{C}{Q}
    \quad \mathrm{ iff } \quad
    ∀ s,ŝ'.\
        P(s) ∧ C(s,ŝ') ⇒ ŝ' ∈ S ∧ Q(ŝ') \]
\end{definition}
Given a starting state~$s$ with $P(s)$, the possible final states~$ŝ'$
after executing command~$C$ satisfy two constraints: 
They must be regular states $ŝ' ∈ S$, ruling out runtime errors in the loop body,
and they must satisfy the postcondition~$Q(ŝ')$ of the triple.
Analogously to splitting total correctness into termination and partial correctness,
we separate these aspects of safe execution and correctly establishing the postcondition into two semantic judgements:
\begin{definition}[Safety and Correctness of Hoare-Triples]
    \label{def:safe}
    \label{def:correct}
\begin{align}
\safe{P}{C}
    \quad &\mathrm{ iff }\quad
    ∀ s,ŝ'.\
        P(s) ∧ C(s,ŝ') ⇒ ŝ' ∈ S
    \\
\correct{P}{C}{Q}
    \quad &\mathrm{ iff }\quad
    ∀ s,ŝ'.\
        P(s) ∧ C(s,ŝ') ∧ ŝ' ∈ S ⇒ Q(ŝ')
\end{align}
\end{definition}
which clearly satisfy this correspondence:
\begin{align}
    \label{eq:split}
⊧ \hoare{P}{C}{Q}
    \quad\textrm{ iff }\quad
    \safe{P}{C}
    \textrm{ and }
    \correct{P}{C}{Q}
\end{align}

\begin{definition}[Semantics of Loops]
    \label{def:loop}
Semantically, a loop $W(t,B) ⊆ S × Ŝ$ with test~$t ∈ S$ and body~$B ⊆ S × Ŝ$ is defined as the least fixpoint of
\begin{align*}
¬ t(s)
    & ⇒ W(t,B)(s,s) \\
t(s) ∧ B(s,\err)
    & ⇒ W(t,B)(s,\err) \\
t(s) ∧ B(s,\brk s')
    & ⇒ W(t,B)(s,s') \\
t(s) ∧ B(s,s') ∧ W(t,B)(s',ŝ'')
    & ⇒ W(t,B)(s,ŝ'')
\end{align*}
\end{definition}
The first condition terminates the loop,
the second condition propagates errors in the body,
the third condition propagates early loop exit,
and the last condition unrolls the loop once if the first iteration results in a regular state $s' ∈ S$.

In the following we are concerned in verifying correctness of a loop $W(t,B)$
wrt. pre-states~$P$ and post-states~$Q$, as expressed by $⊧ \hoare{P}{W(t,B)}{Q}$, respectively its constitutents of ``safe'' and ``correct'' execution via \eqref{eq:split}.

\section{Verification of Loops with Invariants and Contracts}
\label{sec:approach}

This section succinctly states the approach to the verification with invariants and loop contracts, and we prove soundness and completeness theorems.
The results have been mechanized in Isabelle/HOL~\cite{nipkow2002isabelle}.
The presentation here is based on systems of cyclic Horn clauses~\cite{bjorner2015horn}.
Subsequently, we mark those conditions dealing with runtime errors by~$(\dagger)$,
and those dealing with $\BREAK$s by~$(\ddagger)$.

\begin{definition}[Loop Invariants, \citet{hoare1969axiomatic,floyd1993assigning}]
    \label{def:inv}
Predicate~$I ⊆ S$ is an inductive \emph{invariant} of loop $W(t,B)$ wrt.
pre-state described by~$P ⊆ S$, if
\begin{align*}
P(s₀)
    & ⇒ I(s₀) \\
I(s) ∧ t(s) ∧ B(s,s')
    & ⇒ I(s') \quad \text{\normalfont (when $s' ∈ S$)}
\end{align*}
An inductive invariant $I$ is \emph{safe} wrt. executions of the loop body $B$, if
\begin{align*}
I(s) ∧ t(s) ∧ B(s,\err)
    & ⇒ \False \tag{$\dagger$}
\end{align*}
An inductive invariant $I$ is correct wrt. a post-states $Q ⊆ S$, if
\begin{align*}
I(sₙ) ∧ ¬ t(sₙ)
    & ⇒ Q(sₙ) \\[4pt]
I(s) ∧ t(s) ∧ B(s,\brk{sₙ})
    & ⇒ Q(sₙ) \tag{$\ddagger$}
\end{align*}
\end{definition}
The first condition establishes~$I$ initially,
the second propagates~$I$ over a single iteration of the body otherwise.
The third condition~$(\dagger)$ prevents errors in the body.
The last two lines ensure~$Q$ upon regular termination of the loop,
as well as directly after a $\BREAK$.
Note, $Q(s)$ does not necessarily imply $¬ t(s)$,
i.e., we cannot take the negative loop test for granted
if there are non-local exits of the loop by~$\BREAK$.

Relational invariants~$J ⊆ S × S$ 
are sometimes convenient~\cite{mraihi2013invariant},
where $J(s₀,s)$ additionally tracks the state~$s₀$ when the loop was entered first,
which can of course be encoded with auxiliary variables as
$I(s) \coloneqq ∃ s₀.\ P(s₀) ∧ J(s₀,s)$.

\smallskip

It is clear that we have chosen the notions of safe and correct invariants
to mirror precisely the semantic counterparts of safe and correct loops, respectively:
\begin{theorem}[Soundness of Loop Invariants]
For a loop $W(t,B)$,
    \label{thm:inv}
    \begin{itemize}
    \item given a safe invariant~$I$ wrt.~$P$
          then $\safe{P}{W(t,B)}$, and
    \item given a correct invariant~$I$ wrt.~$P$ and $Q$
          then $\correct{P}{W(t,B)}{Q}$
    \end{itemize}
\end{theorem}
\begin{proof}
We prove $I(s) ∧ W(t,B)(s,ŝ') ⇒ ŝ' ∈ S ∧ Q(ŝ')$ (first claim),
resp. $I(s) ∧ W(t,B)(s,ŝ') ∧ ŝ' ∈ S ⇒ Q(ŝ')$ (second claim),
each by induction over the least fixpoint of \cref{def:loop}
using the relevant conditions from~\cref{def:inv}. \qed
\end{proof}

\begin{theorem}[Completeness of Loop Invariants]
    \label{thm:c}
For a loop $W(t,B)$,
    \begin{itemize}
    \item if $\safe{P}{W(t,B)}$ then there exists
          a corresponding safe invariant $I$
    \item if $\correct{P}{W(t,B)}{Q}$, there is
          a corresponding correct invariant $I$
    \item if $⊧ \hoare{P}{W(t,B)}{Q}$, there is
          an invariant~$I$ that is safe \emph{and} correct
    \end{itemize}
\end{theorem}
\begin{proof}
Inductive invariant $λ s.\ ∃ s₀.\ P(s₀) ∧ I^*(s₀,s)$ proves all three claims,
where $I^* ⊆ S × S$ is the strongest relation that characterizes regularly terminating loop iterations, defined as the least fixpoint of:
\begin{align*}
¬ t(s)
     & ⇒ I^*(s,s) \\
t(s) ∧ B(s,s') ∧ I^*(s',s'')
     & ⇒ I^*(s,s'')
\end{align*}
We omit some technical lemmas that connect $I^*$ with $W(t,B)$. \qed
\end{proof}

The result states that
all critical pieces of information (e.g. outcome of loop test~$t$)
and key reasoning features from the underlying induction proof
are reflected somehow in the constraints of \cref{def:inv}.
Of course, reasoning about $I^*$ is by no means easier than a proof using the semantic definition
and the challenge in practice is to find closed-form solutions in a given background theory.

\begin{definition}[Loop Contract]
A (correct) loop contract $I,R$ consists of a loop precondition~$I$ that is a safe invariant,
and a correct summary~$R$ (cf. below).
\end{definition}

The precondition of a loop contract, as discussed previously, is just a safe invariant (cf. \cref{def:inv}).
The summary component of a contract is a relation~$R ⊆ S × S$ that characterizes remaining iterations,
such that $R(s,sₙ)$ holds between any intermediate state~$s$ at loop head and final state~$sₙ$ at loop exit.
\begin{definition}[Loop Summary]
    \label{def:sm}
Relation $R ⊆ S × S$ is a summary of a loop $W(t,B)$ wrt. precondition~$P ⊆ S$, if
\begin{align*}
¬ t(sₙ)
    & ⇒ R(sₙ,sₙ) \\
t(s) ∧ B(s,s') ∧ R(s',sₙ)
    & ⇒ R(s,sₙ) \\[4pt]
t(s) ∧ B(s,\brk{sₙ})
    & ⇒ R(s,{sₙ}) \tag{$\ddagger$}
\end{align*}
A summary~$R$ is called \emph{correct} for a postcondition~$Q ⊆ S$, if
\begin{align*}
P(s₀) ∧ R(s₀,sₙ)
    & ⇒ Q(sₙ)
\end{align*}
\end{definition}
The first line establishes that~$R$ holds reflexively at a regular loop exit,
as the dual of the initialization condition of invariants.
The second line lifts~$R$ of from remaining iterations until termination of the loop to a summary that accounts for an additional leading iteration, whereas
the third line~$(\ddagger)$ establishes that~$R$ summarizes the last partial execution of the loop body upon a $\BREAK$.

The last line applies the relation $R$ to the original pre-state~$s₀$ that satisfies~$P$ to establish~$Q$.
Assumption $P(s₀)$ is the counterpart to the negated loop test $¬ t(sₙ)$ in the exit condition of \cref{def:inv}.
If the loop body $B$ does not contain breaks at all---which can be checked syntactically---
we may enrich~$R(s,sₙ)$ with~$¬ t(sₙ)$ for free, effectively adding it as an assumption in line three.

There is no safe counterpart for summaries,
because they wrap up a loop execution after the fact when it would be too late
to catch runtime errors.

Loop contracts implicitly translate a loop into a tail-recursive procedure.
The summary, taking the role of its postcondition,
can then be interpreted as a relation between the \emph{parameters} of the procedure and its \emph{return value}.
This provides an intuitive justification why non-relational of summaries are not adequate.%
    \footnote{Dually to invariants,
              non-relational version of summary~$R$
              would quantify over final states as
              $∀ s'.\ ¬ t(s') ⇒ R(s,s')$,
              but that condition is too strong at loop exit.}

\begin{theorem}[Soundness of Summaries]
    \label{thm:sm}
Given a correct summary~$R$ of loop $W(t,B)$
that satisfies \cref{def:sm}
wrt.~$P$ and~$Q$,
then $\correct{P}{W(t,B)}{Q}$.
\end{theorem}
\begin{proof}
We prove $W(t,B)(s,ŝ') ∧ ŝ' ∈ S ⇒ R(s,ŝ')$
by induction over the least fixpoint from \cref{def:loop}, the claim follows.
\qed
\end{proof}

Of course, it is possible to strengthen the conditions of \cref{def:sm}
by known inductive invariants, occurring as additional assumptions.
While this is rather convenient in practice,
it is not necessary in theory:

\begin{theorem}[Completeness of Loop Summaries]
    \label{thm:cc}
    For a loop $W(t,B)$,
if $\correct{P}{W(t,B)}{Q}$,
then there exists a corresponding loop summary~$R$.
\end{theorem}
\begin{proof}
We take $R^* ⊆ S × S$, defined as least fixpoint of
\begin{align*}
¬ t(s)
    & ⇒ R^*(s,s) \\
t(s) ∧ B(s,\brk s') ∧ R^*(s₀,s)
    & ⇒ R^*(s₀,s') \\
t(s) ∧ B(s,s') ∧ R^*(s',s'')
    & ⇒ R^*(s,s'')
\end{align*}
This $R^*$ is the strongest relation that characterizes terminating loop iterations, possibly ending with a $\BREAK$ command (in contrast to $I^*$ of \cref{thm:c}).
We rely on the presence of $P(s₀)$ in the third condition of \cref{def:sm}: $R^*(s₀,sₙ)$ implies $W(t,B)(s₀,sₙ)$, which proves $Q(sₙ)$ via validity of the Hoare triple.
    \qed
\end{proof}

\begin{corollary}[Adequacy of Loop Contracts]
$⊧ \hoare{P}{W(t,B)}{Q}$
if and only if there exists a corresponding loop contract~$I,R$ wrt.~$P$ and~$Q$.
\end{corollary}

\section{Translating between the Approaches}
\label{sec:bridge}

Having soundness and completeness of both approaches
from \cref{sec:approach}
we now characterize their relationship.

\begin{corollary}
    \label{cor:syntactic}
For a given loop $W(t,B)$
there exists a safe and correct invariant~$I$ that satisfies \cref{def:inv}
wrt.~$P$ and~$Q$,
if and only if there exists a correct contract $J,R$ such that $J$ is a safe invariant and
$R$ satisfies \cref{def:sm}.
\end{corollary}
\begin{proof}
In both directions, we have that $⊧ \hoare{P}{W(t,B)}{Q}$ by \cref{thm:inv,thm:sm},
respectively, the claim then follows by \cref{thm:c,thm:cc}.
    \qed
\end{proof}

This corollary is of course not surprising,
but the proof via the completeness theorems and the underlying constructions $I^*$ and $R^*$ is unsatisfactory.
A~direct translation that avoids these artifacts is clearly more useful,
by constructing $I$ from $J$ and~$R$, and vice-versa, $R$ from $I$ and $Q$, as
shown with \cref{prop:lift,prop:lower}.
Not only does this give a direct and obvious proof of \cref{cor:syntactic},
it also tells us how to integrate tools for the respective approaches as discussed subsequently.

\begin{proposition}[Invariants from Contracts]
    \label{prop:lift}
Given a contract~$J,R$ that satisfies \cref{def:sm}, then
$I$ is a safe and correct invariant that satisfies \cref{def:inv}, where:
\begin{align}
    \label{eq:lift}
I(s) ~≔~ ∃ s₀.\ P(s₀) ∧ J(s) ∧ \big(∀ sₙ.\ R(s, sₙ) ⇒ R(s₀, sₙ)\big)
\end{align}
\end{proposition}
The first conjunct keeps track of the initial state~$s₀$ that satisfies~$P$,
which is needed to make use of the last property of~$R$ in \cref{def:sm}.
The second conjunct tracks the safe invariant~$J$,
whereas the third conjunct predicts that the loop summary wrt.
the remaining iterations between current state~$s$ and an arbitrary final state~$sₙ$
can be lifted to a summary of the whole execution beginning at $s₀$.

\begin{proof}
We prove conditions of \cref{def:inv} for the lifted invariant~\eqref{eq:lift}.
The interesting part is the choice of $sₙ$ in \eqref{eq:lift},
which is immediate for the loop exit cases with $¬ t(sₙ)$ resp. $\brk{sₙ}$.
Otherwise, $sₙ$ is the same for both instances of~$I$ when propagating it over the iteration of the body.
    \qed
\end{proof}

\Cref{prop:lift} has immediate application in verification tools:
Contracts can be supported straight-forward
in as a \emph{front-end} feature of a deductive verifier like Dafny~\cite{leino2010dafny} that takes specifications from the user.
The only necessary extensions are
adding contract annotations to loops and expressing relational predicates,
e.g., with the widely-used $\old$ keyword or special naming conventions as in VeriFast~\cite{philippaerts2014software}.
The analogue of \eqref{eq:lift} appears to be useful in Separation Logic~\cite{reynolds2002separation}, too.
It has has been noted in a similar form in~\cite{schwerhoff2015lightweight}
as an encoding of Tuerk's approach~\cite{tuerk2010local},
but it is presented less precisely wrt. the states involved
(cf. \cref{sec:related}).

\smallskip

Conversely to \cref{prop:lift}, a tool with first-class support for contracts
can be turned into a purely invariant-based verifier.
The gap between the conditions of \cref{def:inv} and \cref{def:sm}
wrt. condition~$Q$ can be closed by canonical summaries:
\begin{proposition}[Contracts from Invariants]
If $I$ is a safe and correct invariant then $I,R$ is a correct contract,
where
    \label{prop:lower}
\begin{align}
R(s,sₙ) &~≔~ I(s) ⇒ ¬ t(sₙ) ∧ I(sₙ)
    && \text{for loops without } \BREAK \label{eq:lower} \\
R(s,sₙ) &~≔~ I(s) ⇒ Q(sₙ)
    && \text{for all loops, possibly with } \BREAK \tag{$\ddagger$}
\end{align}
\end{proposition}
Intuitively, these summaries characterize that if we jump into the loop
with a state~$s$ that satisfies the invariant~$I(s)$,
then the remaining iterations will establish precisely what can
be derived from \cref{def:inv} for the final state~$sₙ$, respectively.
\begin{proof}
We prove the conditions of \cref{def:sm} for this~$R$ from \cref{def:inv} for~$I$.
Premise $I(s)$ is needed for the respective exit properties of~$I$ to demonstrate $R(sₙ, sₙ)$
when the loop terminates in $s = sₙ$.
Negative polarity of $I(s)$
turns the known forward propagation of~$I$ into
the required backward propagation of~$R$.
    \qed
\end{proof}

While technically correct, the constructions \eqref{eq:lift} and \eqref{eq:lower}
produce rather large and unwieldy formulas.
One might furthermore worry that the introduction of the universal quantifiers into invariant~\eqref{eq:lift} hampers proof automation when~$R$ is a complex formula,
indeed, Dafny would typically fails to infer appropriate triggers for it.
There is a class of properties where the overhead of the respective translation disappears completely.
This is the case when we are interested in tracking a function~$f(x)$ over state variables~$x$ (or analogously a predicate $p(x)$).
\begin{proposition}[Functional Invariants and Summaries]
    \label{prop:functional}
For a loop $W(t,B)$ with precondition~$P$ and program variables~$x,y$:
\begin{itemize}
\item $∃ x₀.\ P(x₀) ∧ f(x) = f(x₀)$ is an invariant iff $f(x') = f(x)$ is a summary
\item $y' = f(x)$ is a summary implies that $f(x) = f(x₀)$ is an invariant
\end{itemize}
merely by simplifying the result of the respective translations~\eqref{eq:lift} and \eqref{eq:lower}. \qed
\end{proposition}
Note, the two cases coincide when $f(x') = y'$ for all final~$x',y'$ with~$¬ t(x',y',\ldots)$.

\smallskip

\noindent \emph{Example:}
Recall the specification of the loop in \code{max} from \cref{fig:max}
using loop precondition
$0 ≤ \code{l} ≤ \code{r} < \code{n}$
and summary
$\code{a}[\code{l}'] = \max(\code{a}[\code{l}..\code{r}+1])$.
Given that finally $\code{l} = \code{r}$, the invariant constructed via \cref{prop:lift} simplifies to:
\begin{align}
    \label{ex:1}
\textbf{invariant} \qquad
0 ≤ \code{l} ≤ \code{r} < \code{n}
   ~∧~ \max(\code{a}[\code{l}..\code{r}+1]) = \max(\code{a}[0..\code{n}])
\end{align}
In \cref{sec:examples} we will see that this simplification applies fairly often in practice and can uncover invariants that are notably different from the textbook solutions,
yet conceptually simple and insightful in some sense.
Simplifying the translation in the converse direction can work nicely, too.
Recovering \emph{reasonable} summaries from the invariants from \cref{sec:motivation}, however, is challenging.
We do not necessarily expect the reader to follow all the details,
but for completeness of discussion we do include an example for the invariant of the KIV team.

\smallskip

\noindent \emph{Example:}
The KIV invariant has the form $I(s) \equiv ∃ k.\ J(k,s)$
where~$J$ consists of two conjuncts, $0 ≤ \code{l} ≤ k ≤ \code{r} < n$ and 
$\code{a}[k] = \max(\code{a}[0..\code{n}])$.
We now construct a summary following \eqref{eq:lower} (first variant),
by simplification steps as well as some heuristic weakening to obtain the original summary shown in \cref{sec:motivation} as \eqref{eq:kiv-summary} below.

Critically, we choose the \emph{same}~$k$ for both occurrences of $I$ in~\eqref{eq:lower}, i.e., yielding $J(k,s) ⇒ ¬ t(sₙ) ∧ J(k,sₙ)$ abstractly,
which makes explicit the functional dependency for $\code{a}[k]$ for that~$k$.
This roughly corresponds to the intuition that such a fixed index~$k$ can be determined (deterministically) upfront.
Moreover, we make use of the following fact
\begin{align*}
\textbf{lemma} \quad
    0 ≤ \code{l} ≤ k ≤ \code{r} < n ⇒
    \max(\code{a}[0..\code{n}]) = \max(\code{a}[l..\code{r}+1])
\end{align*}
to generalize all occurrences of~$\max(\code{a}[0..\code{n}])$,
and we note that the loop terminates with $\code{l}' = k = \code{r'}$,
such that $¬t(sₙ) ∧ I(sₙ)$ simplifies to
\begin{align}
\textbf{summary} \quad
    \label{eq:kiv-summary}
    0 ≤ \code{l}' = \code{r}' < n ∧ \code{a}[\code{l}'] = \max(\code{a}[l..\code{r}+1])
\end{align}

\section{Loop Contracts in Hoare Logic}
\label{sec:hoare}

In this section we present a Hoare logic proof rule
for the verification of loops with contracts
that lends itself to a straight-forward implementation.
The approach works analogously with strongest postcondition and weakest precondition predicate transformers,
as explained in Alexandru's thesis~\cite{alexandru2019}.
The idea mirrors that of \citet{tuerk2010local}, however,
he uses a shallow embedding of formulas, programs, and Hoare triples in the higher-order logic of the HOL system.
The inductive case for one iteration of the body~$B$ with test~$t$ is expressed in~\cite{tuerk2010local} as
\begin{align}
    \label{eq:tuerk}
∀\ x,C.\
    \big(∀\ y.\ \hoare{P(y)}{C}{Q(y)}\big)
        ⇒ \hoare{t(x) ∧ P(x)}{B; C}{Q(x)}
\end{align}
where the premise of the implication amounts to the inductive hypothesis
for the remaining loop iterations, abstracted here by arbitrary the command~$C$.
The primary question is how to represent the inductive hypothesis in \eqref{eq:tuerk}
without escaping to the meta-level with Hoare triples as first-class objects.
The key insight is that \emph{specification statements}~(\citet{morgan1988specification})
lead to an elegant formulation as rule \textsc{LoopContract} below.

To this end, we make the distinction between syntax and semantics more precise:
Predicates $P,Q,I,R$ are represented as formulas here,
where a relational summaries~$R$ may refer to primed variables as in \cref{sec:motivation}.
By $P[\xs ↦ \ys]$ we denote the parallel renaming of variables $\xs$ to $\ys$ in $P$,
where by convention we overline vectors of variables~$\xs$.
Derivability of Hoare triples is written $⊦ \hoare{P}{C}{Q}$.
We omit treatment of $\BREAK$ for simplicity, see e.g.~\cite{huisman2000java}.

A specification statement $\SPEC{\xs}{P}{Q}$ has
a precondition~$P$, a set of variables~$\xs$ that are nondeterministically modified, and a relational postcondition~$Q$ that constrains the transition.
The proof rule for the specification introduces new logical variables
that capture the pre-state, and removes the primes in~$Q$:
\begin{align*}
    \label{rule:spec}
\infer[\textsc{Spec}]
    {⊦ \hoare{P}
                  {~\SPEC{\xs}{P}{Q}~}
                  {Q[\xs,\xs'↦\xs_0,\xs]}}
    {\xs_0 \text{ fresh}}
\end{align*}

The proof rule to verify a loop $\WHILE\ t\ \DO\ B$
with test~$t$ and body~$B$ using a contract $I,R$ is show below.
The variables~$\xs = \text{mod}(B)$ are those modified by body~$B$,
and $\xs_0$, $\xs_i$, $\xs_n$ are fresh logical variables
capturing intermediate states.
\begin{align*}
\infer[\textsc{LoopContract}]
    {⊦ \hoare{P}
                  {~\WHILE\ t\ \DO\ B~}
                  {Q}}
    {\begin{array}{c}
        ⊦ \hoare{P}{~\SPEC{\xs}{I}{R}~}{Q} \\[2pt]
        ⊦ \hoare{I ∧ ¬ t ∧ \xs = \xs_n}
                     {~\SKIP~}
                     {R[\xs,\xs'↦\xs_n,\xs]} \\[2pt]
        ⊦ \hoare{I ∧ t ∧ \xs = \xs_i}
                     {~B;\ \SPEC{\xs}{I}{R}~}
                     {R[\xs,\xs'↦\xs_i,\xs]}
     \end{array}}
\end{align*}
The first premise abstracts the computation of the entire loop
in terms of its contract~$I,R$ using a specification statement.
From rule \textsc{Spec}, proof obligations akin to those in \cref{def:sm} are immediate via an application of the consequence rule.
The second premise terminates the loop when~$¬ t$,
the variables~$\xs_n$ capture this final state.
We encode the corresponding proof obligation as a Hoare triple with command~$\SKIP$,%
    \footnote{\citet{tuerk2010local} remarks that, more generally,
              the inductive hypothesis may encompass a subsequent program fragment~$C$
              right after the loop, i.e.,~$\WHILE\ t\ \DO\ B; C$,
              and this (concrete) $C$ would then replace~$\SKIP$ in the second premise, with $\xs = \textrm{mod}(B,C)$.}
which is equivalent to $I ∧ ¬ t ⇒ R[\xs,\xs'↦\xs,\xs]$,
mirroring the exit condition from \cref{def:sm}.
The third premise, in contrast to~\eqref{eq:tuerk},
embeds the inductive hypothesis directly
 into the program as specification statement~$\SPEC{\xs}{I}{R}$ that summarizes the remaining iterations after executing~$B$ once.
Variables~$\xs_i$ capture the state right before~$B$ is executed for later reference in the postcondition,
whereas the reference to the intermediate state $i+1$ after~$B$,
needed for~$R$ in~$\SPEC{\xs}{I}{R}$, is handled implictly via rule~\textsc{Spec}.
Thus, rule \textsc{LoopContract} nicely retains the syntax-oriented, compositional nature of Hoare's approach.

\section{Specification of Examples and Comparison}
\label{sec:examples}

In this section we specify some verification challenges using invariants and contracts.
The examples are chosen not for their difficulty,
but because they highlight specific aspects, advantages,
and limitations of the respective approaches.

Throughout, we employ the $\old$ keyword in loop preconditions resp. invariants
to refer back to the state before the loop.
Variables in procedure annotations always refer back
to the initial values of the parameters
and $\mathbf{res}$ denotes the result value returned by the procedure.
Moreover, we factor out the part of the verification that is common to both approaches,
in terms of a loop precondition that one should understand as part of the
invariant as well as the loop contract.
Finally, for those algorithms that do not contain break,
we implicitly assume that the negated loop test is made part of the summary
(cf. \cref{def:sm} and \cref{prop:lower}).

\paragraph{Fast Exponentiation.}
The fast exponentiation algorithm computes $\code{x}^\code{n}$
by traversing over the binary representation of the exponent \code{n}.
The program%
    \footnote{Presentation adapted from \url{http://toccata.lri.fr/gallery/power.en.html}}
tracks a multiplier $2^\code{p}$ for each binary digit
that is applied to the intermediate result \code{r}
only if a binary digit is set to one,
where residual exponent \code{e} continuously shifts right such that the lowest significant
binary digit of \code{e} always corresponds to \code{p}.

\begin{minipage}{0.45\textwidth}
\begin{lstlisting}
int fastexp(int x, int n)
  requires $0 ≤ \code{n}$
  ensures  $\mathbf{res} = \code{x}^\code{n}$
{
  int r = 1, p = x, e = n;
  while(e > 0) {
    if(e%2 == 1)
      r = r * p;
    p = p * p;
    e = e / 2;
  }
  return r;
}
\end{lstlisting}
\end{minipage}
    \quad
\begin{minipage}{0.5\textwidth}
    \textbf{precondition (both)} \\[2pt]
    $0 ≤ \code{e}$

    \medskip

    \textbf{invariant} \\[2pt]
    $\code{r} · \code{p}^\code{e} = \code{x}^\code{n}$

    \medskip

    \textbf{summary} \\[2pt]
    $\code{r}' = \code{r} · \code{p}^\code{e}$
\end{minipage}

Both the invariant as well as the summary
require the same kind of generalization,
to account for the the intermediate result in variable \code{r}.
This example admits a functional characterization according to \cref{prop:functional}.
The route from the summary to the invariant is straight-forward,
the converse can be best understood by noting that the initial values of $\code{r}$, $\code{p}$, $\code{e}$
denoted $\code{r}_0$, $\code{p}_0$, $\code{e}_0$ as in \cref{sec:motivation},
coincide with $1$, $\code{x}$, and $\code{n}$,
such that the invariant can be written as 
    $\code{r} · \code{p}^\code{e} = \code{r}_0 · \code{p}_0^{\code{e}_0}$.

\paragraph{Linear Search.}
Linear search, as shown below, traverses an array~\code{a}
of length~\code{n} from front to back
using index~\code{i} to find an element~\code{x}.
To avoid using \code{return} inside the loop (which we have not formalized in \cref{sec:approach})
we maintain a variable~\code{r} that becomes true once the element is found,
in which case we \code{break} out of the loop.
Remember that in this case we establish the postcondition of the procedure directly,
such that $\lnot \code{r}$ is a valid invariant.

\begin{minipage}{0.5\textwidth}
\begin{lstlisting}
bool lsearch(int x, int a[], int n)
  requires $0 ≤ \code{n}$
  ensures  $\mathbf{res} \iff \code{x} ∈ \code{a}[0..\code{n}]$
{
  int i = 0; bool r = false;
  while(i < n) {
    if(x == a[i])
      { r = true; break; }
    i++;
  }
  return r;
}
\end{lstlisting}
\end{minipage}
    \quad
\begin{minipage}{0.5\textwidth}
    \textbf{precondition (both)} \\[2pt]
    $0 ≤ \code{i} ≤ \code{n} ∧ ¬ \code{r}$

    \medskip

    \textbf{invariant} \\[2pt]
    $\code{x} ∉ \code{a}[0..\code{i}]$

    \medskip

    \textbf{summary} \\[2pt]
    $\code{r}' ⇔ \code{x} ∈ \code{a}[\code{i}..\code{n}]$

    \medskip

    \textbf{invariant} via \cref{prop:lift} \\[2pt]
    $\code{x} ∈ \code{a}[\code{i}..\code{n}] ⇔ \code{x} ∈ \code{a}[0..\code{n}]$
\end{minipage}

The common condition is about the range of the index variable \code{i}
and the fact that the loop head is encountered only with when \code{r} is false.
The invariant states, as expected, that the element has not been found yet
in the initial range up to and not including \code{i}.
The loop postcondition states that the final value of $\code{r}$,
denoted $\code{r}'$,
will indicate whether the element is found in the remaining range between \code{i} and \code{n}.
It is quite similar to the procedure contract of \code{lsearch},
requiring only the generalization of the lower bound.
Moreover, given $¬ \code{r}$, if we write the invariant equivalently as
    $\code{r} ⇔ \code{x} ∈ \code{a}[0..\code{i}]$,
then the respective approaches become entirely symmetric.
The loop invariant lifted from the postcondition via \cref{prop:lift}
gives a nice alternative characterization of the work that remains to be done
(searching from~\code{i}) in relation to the overall work to be achieved.

\paragraph{Binary Search.}
In contrast to linear search, binary search tracks two indices,
somewhat similarly to maximum by elimination from \cref{sec:motivation}.
The code is shown below, using lower index \code{l}
and upper index \code{u} (both inclusive).

\begin{minipage}{0.5\textwidth}
\begin{lstlisting}
bool bsearch(int x, int a[], int n)
  requires $0 ≤ \code{n} ∧ \sorted(a)$
  ensures  $\mathbf{res} \iff \code{x} ∈ \code{a}[0..\code{n}]$
{
  int l = 0, u = n-1;
  bool r = false;
  while(i < n) {
    int m = (l+u) / 2;
    if (x > a[m])      { l = m+1; }
    else if (x < a[m]) { u = m-1; }
    else       { r = true; break; }
  }
  return r;
}
\end{lstlisting}
\end{minipage}
    \quad
\begin{minipage}{0.5\textwidth}
    \textbf{precondition (both)} \\[2pt]
    $0 ≤ \code{i} ≤ \code{n} ∧ ¬ \code{r}$

    \medskip

    \textbf{invariant} \\[2pt]
    $\code{x} ∉ \code{a}[0..\code{l}]
     ∧ \code{x} ∉ \code{a}[\code{u}+1..\code{n}]$

    \medskip

    \textbf{summary} \\[2pt]
    $\code{r}' ⇔ \code{x} ∈ \code{a}[\code{l}..\code{u}+1]$

    \medskip

    \textbf{invariant} via \cref{prop:lift} \\[2pt]
    $\code{x} ∈ \code{a}[\code{l}..\code{u}+1] ⇔ \code{x} ∈ \code{a}[0..\code{n}]$
\end{minipage}

The invariant for binary search now excludes two sub-ranges of the array,
whereas the summary incorporates only the minor additional generalization
for the upper bound from \code{n} to $\code{u}+1$.
Like in \code{max} from \cref{sec:motivation} but unlike with linear search,
the array is divided into three logical parts,
of which the shown invariant considers two,
whereas summaries can zoom into the single remaining part.
This effect has been noted by \citet[Sec.~2.3]{furia2010inferring}
where it is addressed by a heuristic called ``uncoupling'' that splits up ranges as needed.
With respect to binary search and similarly maximum by elimination from \cref{sec:motivation},
approaching the problem via contracts leads to a nice invariant
via \cref{prop:lift} that avoids such uncoupling.

\paragraph{Phone Number Comparison.}
Summaries can mediate between a forward computation,
which is effectively a ``left-fold'', and a correctness condition that is a ``right-fold'' (cf. \cite{hutton1999tutorial}).
This case occurs e.g. when the logical specification uses an intermediate abstraction step
to algebraic lists or sequences, over which functions and predicates
are typically specified by structural recursion.

\newcommand{\nil}{\code{nil}}
\newcommand{\cons}{\code{cons}}
\newcommand{\head}{\code{head}}
\newcommand{\tail}{\code{tail}}
\newcommand{\reverse}{\mathit{reverse}}
\newcommand{\append}{\mathit{append}}
\newcommand{\filter}{\mathit{filter}}

Consider the comparison of phone numbers%
    \footnote{Example communicated by Rustan Leino, who based his verification on~\cref{inv:phone}..}
by ignoring non-digit characters.
As an example, the phone numbers \code{(0) 12/345} and \code{01-2345} should be regarded the same, whereas \code{1-23-45} is different because of the missing leading~\code{0}.

\begin{minipage}{\textwidth}
\begin{lstlisting}
bool compare(int a[], int m, int b[], int n)
  ensures  $\mathbf{res} \iff \filter(\code{isdigit},\code{a}[0..\code{m}]) = \filter(\code{isdigit},\code{b}[0..\code{n}])$
{
  bool r = false;
  int i = 0, j = 0;
  while(true) {
    if(i == m && j == n)                    { r = true; break; }
    else if(i < m && !isdigit(a[i])         { i++; }
    else if(j < n && !isdigit(b[j])         { j++; }
    else if(i < m && j < n && a[i] == b[j]) { i++; j++; }
    else                                    { r = false; break; }
  }
  return r;
}
\end{lstlisting}
\end{minipage}

The algorithm keeps two indices, $\code{i}$~and~$\code{j}$, into arrays~$\code{a}$ and~$\code{b}$
that store the characters of the respective numbers, of lengths~$\code{m}$ and $\code{n}$.
The algorithm consists of a loop that increments~$\code{i}$ and~$\code{j}$
according to several cases until the numbers are fully compared
(first \code{if}) or a mismatch is detected (last \code{else}).
If~$\code{a}[\code{i}]$ is not a digit then $\code{i}$ moves forwards, similarly~$\code{j}$ for~$\code{b}[\code{j}]$,
and both jointly move forward over two equal digits.
The result of the comparison is returned via variable $\code{r}$ as previously.

A nice specification of this algorithm is in terms of algebraic lists,
constructed from $\nil$ and $\cons$,
where $a[i..j]$ denotes $\cons(a[i], \cons(\ldots, \cons(a[j-1], \nil)))$,
the list of the elements from index~$i$ to and including~$j-1$ of~$a$.
We rely on a function $\filter$
that keeps only those elements in the list that satisfying a predicate~$p$.
Filtering is defined by structural recursion over the algebraic list:
\begin{align*}
\filter(p,\nil) &= \nil
    \\
\filter(p,\code{cons}(x,xs))
    &= \begin{cases}
       \cons(x, \filter(p,xs)), & \text{if } p(x) \\
       \filter(p,xs)),          & \text{otherwise.}
       \end{cases}
\end{align*}

With these prerequisites, the specification of procedure \code{compare}
states that the boolean result indicates whether filtering for digits only produces identical lists.
An obvious candidate for the invariant follows the idea
from linear search, and generalizes the upper bounds of the ranges compared from the array
length to the respective counter variable:
\[ \text{{\bf invariant} (problematic)}
\quad \filter(\code{isdigit},\code{a}[0..\code{i}]) = \filter(\code{isdigit},\code{b}[0..\code{j}]) \]
While correct, the approach has the significant drawback that we need a lemma that
unfolds the recurrence of $\filter(p,\code{a}[0..k+1]$) \emph{at the end} instead of the front like the definition,
to accommodate the index increments, which in turn is provable by induction only after a further generalization
of the lower index from~0 to a variable.
Overall, this approach is somewhat cumbersome,
and can be avoided when the solution is approached with loop contracts:
\[ \textbf{summary} 
\quad \code{r}' \iff \filter(\code{isdigit},\code{a}[\code{i}..\code{m}]) = \filter(\code{isdigit},\code{b}[\code{j}..\code{n}]) \]
for which the proof is straight-forward,
and from which we immediately get an equally easy to prove invariant by \cref{prop:lift}:
\begin{align}
    \label{inv:phone}
\textbf{invariant} \hspace{1cm}
          & \filter(\code{isdigit},\code{a}[0..\code{m}])
            = \filter(\code{isdigit},\code{b}[0..\code{n}]) \\
{} \iff\ & \filter(\code{isdigit},\code{a}[\code{i}..\code{m}])
            = \filter(\code{isdigit},\code{b}[\code{j}..\code{n}]) \nonumber
\end{align}
The shown mismatch between the natural direction of the loop vs. that of the property
is correlates to the insight that for some algorithms, a recursive version is easier to verify, and some tools explicitly translate
loops into recursive procedures to that end~\cite{blanc2013overview}.
Using loop contracts, one can avoid this intermediate step.
For a similar discussion in the context of separation logic we refer to
\cite{tuerk2010local,huisman2015verifythis}.

We emphasize that it is not always possible to factor out specification functions like $\filter$ nicely and to defer the complexity of additional lemmas to a library,
as one might be inclined to suggest.
When such functions are specific to the case study,
or simply when higher order functions are not supported by the tool,
the ability to base the loop specification on right-fold loop summaries
is certainly a useful trick in the bag that deserves to be treated first class.

\paragraph{Array Copy.}
We turn to programs that manipulate arrays, which uncovers a deficit of summaries.
Recall that loop contracts reason about \emph{three} states,
an initial one $s₀$, an intermediate one $s$, and a final one~$sₙ$
(cf. \cref{fig:sm}),
whereas invariants reason only about the first \emph{two}.
While it increases expressive power as demonstrated above, it comes at a cost, too:
Summarizing the remaining loop iterations from~$s$ to the final state~$sₙ$
by summary~$R(s,sₙ)$ does not automatically reflect the
array modifications applied to get from~$s₀$ to~$s$.
This occurs in the program below that copies~\code{n}
entries from array \code{a} to \code{b}.
This is an instance of the more general problem of \emph{framing},
that is long known~\cite{burstall1972some} and well studied.
A comprehensive treatment is beyond the scope of this paper,
but we show how it surfaces in the example.

\begin{minipage}{0.5\textwidth}
\begin{lstlisting}
void copy(int a[], int b[], int n)
  requires $0 ≤ \code{n}$
  ensures  $\code{b}[0..\code{n}] = \code{a}[0..\code{n}]$
{
  int i = 0;
  while(i < n) {
    b[i] = a[i];
    i = i+1;
  }
}
\end{lstlisting}
\end{minipage}
    \quad
\begin{minipage}{0.5\textwidth}
    \textbf{precondition (both)} \\[2pt]
    $0 ≤ \code{i} ≤ \code{n}$

    \medskip

    \textbf{invariant} \\[2pt]
    $\code{b}[0..\code{i}] = \code{a}[0..\code{i}]$

    \medskip

    \textbf{summary} \\[2pt]
    $\code{b}'[0..\code{i}] = \code{b}[0..\code{i}]$ \\
    $\code{b}'[\code{i}..\code{n}] = \code{a}[\code{i}..\code{n}]$
\end{minipage}

The expected invariant specifies that the prefix up to current index \code{i}
has been copied already.
Analogously, the summary predicts that executing the remainder of the loop
will copy the suffix starting from \code{i} into the resulting array~$\code{b}'$
(second equation in the listing).
However, that is not enough:
In the back-propagation step from $\code{i}+1$ to $\code{i}$,
from $\code{b}'[\code{i}+1..\code{n}] = \code{a}[\code{i}+1..\code{n}]$
alone we cannot conclude
$\code{b}'[\code{i}..\code{n}] = \code{a}[\code{i}..\code{n}]$
because the assignment to the entry $\code{b}[\code{i}]$ in the current entry
does not appear anywhere, it is ``forgotten''.
The condition missing from the summary is that the remaining iterations
do not touch the indices again that were modified so far,
i.e., that the lower range of $\code{b}$ is unmodified.

\paragraph{Bubble Sort.}
We now turn to sorting algorithms, which are a classic example, starting with bubble sort.
It turned out to be somewhat tricky to get out of the mindset associated with invariants,
and to find a nice notation for a natural specification.
Again, framing is crucial, and we will do it explicitly, in terms of comparing array ranges $a[i..j] = b[i..j]$.
In similar spirit, by $a[i..j] \rightleftharpoons b[i..j]$ we denote that the array range $a[i..j]$ is a permuation of $b[i..j]$, which is equivalent to stating that the multiset
of elements on both sides are the same (a common encoding in Dafny).
The code is shown below, together with a graphical visualization.

\begin{minipage}{0.5\textwidth}
\begin{lstlisting}
void bubblesort(int a[], int n)
  requires $0 ≤ \code{n}$
  ensures  $\code{a} \rightleftharpoons \old(\code{a})$
  ensures  $\mathit{sorted}(\code{a})$
{
  for(int i = n; i > 1; i--)
    for(int j = 0; j < i-1; j++)
      if(a[j] > a[j+1])
        swap(j, j+1, a);
}
\end{lstlisting}
\end{minipage}
\begin{minipage}{0.5\textwidth}
\includegraphics[width=4cm]{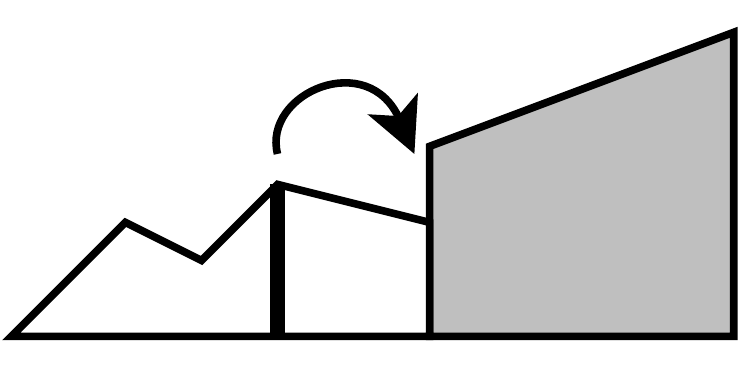}
\end{minipage}

The algorithm gradually constructs a sorted suffix of the array,
which is shaded in grey in the figure on the right.
Goal of the inner loop is to move the largest element in the prefix
up to the boundary of the sorted range, as visualized by the arrow.
The loop specifications are shown below.

\newcommand{\eqn}{\mathrel{\makebox[\widthof{$\rightleftharpoons$}]{$=$}}}

\begin{minipage}[t]{0.5\textwidth}
\underline{Outer Loop}

\medskip

\textbf{precondition (both)}

$0 ≤ \code{i} < \code{n}$

    \medskip

\textbf{invariant}

$\code{a} \rightleftharpoons \old(\code{a})$

$\mathit{max}(\code{a}[0..\code{i}]) ≤ \mathit{max}(\code{a}[\code{i}..\code{n}])$

$\mathit{sorted}(\code{a}[\code{i}..\code{n}])$

\medskip

\textbf{summary}

$\code{a}'[0..\code{i}] \rightleftharpoons \code{a}[0..\code{i}]$

$\code{a}'[\code{i}..\code{n}] \eqn \code{a}[\code{i}..\code{n}]$

$\mathit{sorted}(\code{a}'[0..\code{i}])$
\end{minipage}
\begin{minipage}[t]{0.5\textwidth}
\underline{Inner Loop}
\medskip

\textbf{precondition (both)}

$0 ≤ \code{j} < \code{\code{i}} ≤ \code{n}$

    \medskip

\textbf{invariant}

$\code{a}[0..\code{i}] \rightleftharpoons \old(\code{a}[0..\code{i}])$

$\code{a}[\code{i}..\code{n}] \eqn \old(\code{a}[\code{i}..\code{n}])$

$0 < j ⇒ \code{a}[\code{j}] = \mathit{max}(\code{a}[0..\code{j}])$

\medskip

\textbf{summary}

$\code{a}'[0..\code{j}] \eqn \code{a}[0..\code{j}]$

$\code{a}'[\code{j}..\code{i}] \rightleftharpoons \code{a}[\code{j}..\code{i}]$

$\code{a}'[\code{i}..\code{n}] \eqn \code{a}[\code{i}..\code{n}]$

$\code{a}'[\code{i}-1] = \mathit{max}(\code{a}'[\code{j}..\code{i}])$

\end{minipage}

In both approaches, we keep track of which parts of the array
have been permuted and which are unchanged, albeit the 
summary is more precise for both loops to account for the frame problem
already seen with array copy.

The approaches for both loops are typically symmetric:
as already seen with linear search, the invariant refers to properties of
the prefix whereas the summary refers to properties of the suffix.
For the inner loop it suffices to keep track of where we have placed the maximum of that range.
For the outer loop, we establish that a particular part is already sorted.

Interestingly, in the contract approach it is not necessary to specify
that all elements in the prefix are smaller than those of the suffix.
This information follows from the strong framing of the outer loop,
together with the summary of the inner loop.

\bigskip

\paragraph{Summary.}
The examples shown in this section complement 
similar expositions of verified algorithms where invariants are used,
for example~\cite{nipkow2020verified,furia2014loop} as well as the Toccata Gallery.%
    \footnote{\url{https://toccata.lri.fr/gallery}}
The intention was to compare utility of loop contracts.
The case is not entirely clear but it is possible to distill some insights
as evidence that sometimes loop contracts are just the right tool.

Contracts reason about \emph{complete} results of a computation,
whereas invariants reason about \emph{partial} intermediate results.
Sometimes, the former is easier to describe, and moreover closer
to the overall correctness property,
which can then be taken from the program's annotation.
This comment applies specifically to those algorithms,
where the work done so far affects the overall result only in minor ways,
as it is the case with the search algorithms,
but not with the ones that modify the array.
We envision that there is a hidden potential to be unlocked
to discover loop specifications automatically, as it is done with invariants in e.g.~\cite{furia2014loop},
and the preliminary experiment in~\cref{sec:evaluation} is a first step into this direction.

Moreover, proof arguments for summaries run counter to
the computation of the loop, implicitly turning it into a recursion.
This helps to bridge the gap when the specification is naturally expressed
as a right-fold.

Overall, expressing correctness properties as part of a summary
provides an alternative and important conceptual angle,
regardless of the underlying verification method.
However, with loop contracts it may be necessary to preserve
some additional information across the round-trip to the final state,
as shown with \code{copy}.
In general, if correctness of the loop depends on unbounded work done so far,
the immediate constraints from executing the body once are insufficient.
Thus, loop precondition resp. the invariant
remain an important conceptual ingredient for formal verification of loops as well.

\section{Evaluation of Proof Automation}
\label{sec:evaluation}

So far, we have emphasized the properties of loop specifications,
how to come up with them from the perspective of a human engineer,
in the context of verification against strong properties,
where full proof automation cannot be expected.

In this section we switch perspective and investigate
how well state-of-the-art solvers for systems of Horn clauses
can be used to infer loop contracts with pre- and postconditions.
The benchmark set of the yearly CHC competition~\cite{rummer2020competition}
contains verification tasks that are based on invariants only,
and consequentially, one might assume that the implemented techniques are geared towards
the shape of the resulting verification conditions.
On the other hand, completeness \cref{thm:cc} of loop contracts suggests,
that we can expect reasonable results.
Furthermore, the verification tasks that are currently in reach of Horn solvers
focus primarily on integer arithmetic, for which effective algorithms exist.
Nevertheless, solvers are forced to encode correctness
primarily into loop postconditions instead.

\paragraph{Research Question:}
\emph{How well do existing solvers instantiate loop contracts automatically in comparison to invariants?}

This question is relevant in so far, as any complex verification task
comprises smaller sub-problems, often numerical, that \emph{can} be automated well,
whereas for other aspects, human guidance is needed.
Tools like Boogie~\cite{leino2008boogie} already implement such features.
In order to make use of loop contracts in practice, it is therefore useful to know
whether such an interplay between automation and manual proof works in that setting, too.

The evaluation is based on a prototype tool
that translates~C programs into systems of Horn clauses like those in \cref{sec:approach}.
The tool implements verification with both invariants and contracts
for a fair comparison.
As benchmark, we consider a those tasks of the SV-COMP repository
which are currently supported by the tool,
and which contain tasks with loops.
We emphasize that it is \emph{not} a goal to improve the overall performance
over the state-of-the-art~\cite{beyer2020advances} across a wide range of benchmarks,
as it would require significant additional engineering effort to support all the C features present in the SV-COMP benchmarks.

\paragraph{Experimental Setup.}
We focus on four categories
from the SV-COMP benchmark repository, \code{Arrays}, \code{Loops}, \code{Recursive}, and \code{ControlFlow},
where~459 of a total of~925 benchmarks are currently supported by our translator.
These benchmarks contain nondeterministic choice,
as well as linear and nonlinear integer arithmetic.
Some tasks, notably those in the \code{Arrays} category,
make use of arrays but pointer aliasing is very limited.
This class of tasks falls generally within the capabilities of the solvers,
albeit most are currently out of reach,
e.g. complex properties over arrays that require quantified invariants,
or when procedure summaries would need to be recursive.

The verification tool used is called \Korn, which is publicly available.
    \footnote{\url{https://github.com/gernst/korn}}
The results presented here have been obtained mid-2020,
\Korn subsequently participated in SV-COMP~2021,
    \footnote{\url{https://sv-comp.sosy-lab.org/2021/results/results-verified/}}
using some additional techniques like random fuzzing,
such that those results are not directly comparable.

We compare Z3~4.8.9 \cite{komuravelli2016smt}
and Eldarica~2.0.4 \cite{hojjat2018eldarica}
as backends,
the top two performers in the CHC competition 2020~\cite{rummer2020competition}.
The hardware configuration is Intel Xeon E3-1230 with 3.40 GHz and 8~cores,
limited to 1 core, 15 min, and 15 GB memory per individual task.
The experiments have been executed on Ubuntu 20.04 with Java~11 (for Eldarica).

\paragraph{Results.}
Numeric results are shown in \cref{tab:results}.
For reference, we include official results%
    \footnote{\url{https://zenodo.org/record/3630205}}
from SV-COMP~2020 for
CPA-seq~\cite{dangl2015cpachecker} (``CPA$\checkmark$''),
Ultimate Automizer~\cite{heizmann2018ultimate} (``UA''),
and VeriAbs~\cite{afzal2020veriabs},
for the same set of 459~tasks (cf. ``\#'').

Both our backend solvers Eldarica and Z3 have trouble with the \code{Array} category,
which is difficult and subject to active research~\cite{deangelis2014verifying,fedyukovich2019quantified}.
VeriAbs~\cite{afzal2020veriabs} performs well,
because it can exploit loop parallelism.
Eldarica can deal quite well with contracts,
and instantiate these even if we force the proof with these,
missing out on 5~tasks wrt. invariants only,
whereas Z3 performs worse with contracts.

\begin{table}[t]
    \centering
    \caption{Correct results, as found by the backend Horn solvers,
             with 15 min CPU time and 15 MB memory, per category.
             Some official results from SV-COMP~2020 are included for reference.
             \#~shows the number of tasks which are supported
             by this evaluation
             vs. the total number of tasks in the category.
             }
    \label{tab:results}
    \setlength{\tabcolsep}{6pt}
    \smallskip

    \begin{tabular}{lrrrrrrrr}
    \toprule
      Category
    & 
    & \multicolumn{4}{c}{Solved Tasks (this paper)}
    & \multicolumn{3}{c}{SV-COMP~2020}
    \\
    \cmidrule(l){3-4}
    \cmidrule(l){3-6}
    \cmidrule(l){7-9}
    &
    & \multicolumn{2}{c}{Eld. 2.0.4~\cite{hojjat2018eldarica}}
    & \multicolumn{2}{c}{Z3 4.8.9~\cite{gurfinkel2016smt}}
    & \multirow{2}{*}{CPA$\checkmark$}
    & \multirow{2}{*}{UA}
    & \multirow{2}{*}{VeriAbs}
    \\
    &   \#/total
    &   inv.
    &   contr.
    &   inv.
    &   contr.
    \\
    \cmidrule(){1-1}
    \cmidrule(lr){2-2}
    \cmidrule(l){3-4}
    \cmidrule(l){5-6}
    \cmidrule(l){7-9}
    Arrays      &  129/436 &   2 &   2 &   7 &   2 &   3 &   3 & 122 \\
    ControlFlow &   39/ 95 &  36 &  36 &  31 &  33 &  37 &  37 &  31 \\
    Loops       &  196/289 & 117 & 112 & 105 &  75 &  83 & 113 & 162 \\
    Recursive   &   95/105 &  82 &  82 &  91 &  91 &  77 &  74 &  83 \\
    \cmidrule(){1-1}
    \cmidrule(lr){2-2}
    \cmidrule(l){2-2}
    \cmidrule(l){3-4}
    \cmidrule(l){5-6}
    \cmidrule(l){7-9}
    Sum         &  459/925 & 237 & 232 & 234 & 201 & 200 & 227 & 398 \\
    \bottomrule
    \end{tabular}
\end{table}

\begin{SCfigure}
    \centering
    \includegraphics[width=0.8\textwidth]{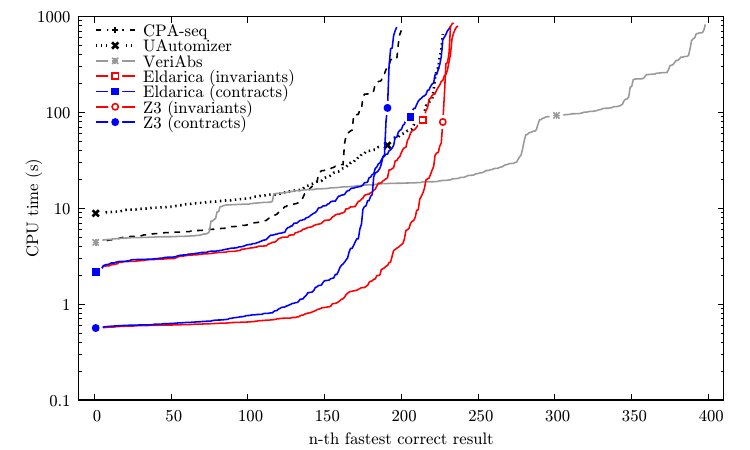}
    \caption{Quantile plot showing the relative performance of the three approaches,
             in comparison to three top software verifiers on the same tasks.}
    \label{fig:quantile}
\end{SCfigure}

\Cref{fig:quantile} show a quantile plot of all correct results,
which in addition to \cref{tab:results} reflects the solving times.
The results for invariants are shown in red,
those with contracts only are shown in blue.
Z3 solves the majority of tasks well within a second,
requiring more than ten seconds only on the last few benchmarks.
We also see that for the configurations using contracts,
if successful, Z3 still takes longer on average.
Eldarica, on the other hand, shows no such clear performance difference
between the configurations.
There are a few incorrect results per experimental setup
(1 for Eldarica, and~2 resp.~3 for Z3),
due to limitations of the implementation,
such as assuming unbounded integers.

\paragraph{Discussion}
We can answer the research question outlined in the beginning positively:
Current Horn clause solvers are in fact able to instantiate contracts,
notably the relational postcondition component,
and their performance is roughly comparable to verification with invariants.

Nevertheless, it should be clear that this evaluation is
not a conclusive assessment of the general utility of loop contracts
for proof automation.
For that, the verification tasks are not entirely suitable.
In particular, integer arithmetic can effectively be dealt with in both settings with established techniques.
On the other hand, consider the category \code{Arrays},
where we could hope that contracts might be beneficial.
However, thes tasks are rather tricky, and would often require quantified invariants,
for which techniques are not mature yet~\cite{bjorner2013solving,fedyukovich2019quantified}

\section{Related Work}
\label{sec:related}

We emphasize that the approach of using contracts, see e.g.~\citet{hehner2005specified} for practical examples.
The work closest to the theory of \cref{sec:approach} is by \citet{hehner1999refinement},
which shows an analogue of \cref{def:sm} as Rule~F in~\cite[Sec 10]{hehner1999refinement}.
Their presentation is closely tied to reasoning about \code{for}-loops,
where starting and ending indices are known (as symbolic expressions).
In contrast, we delimit the loop in terms of precondition~$P$ and postcondition~$Q$,
which lifts the idea to all loops in general.
Moreover, Rule~F does not make explicit the relational nature of postconditions,
whereas Rule~G in~\cite{hehner1999refinement}
considers \emph{two arbitrary} execution segments of the loop, instead of a single segment known to end in a final state (cf. \cref{fig:inv,fig:sm}),
possibly with additional limitations wrt. framing.
Further work that is based on the same idea is~\cite{bohorquez2010elementary,chargueraud2010characteristic}.
A constructive translation like the one shown \cref{sec:bridge}
is not provided there or elsewhere as far as we know.

Ideas to leverage the postcondition of a procedure contract
to derive invariants has been explored by \citet{furia2010inferring}.
Applying such techniques in a setting with loop contracts
is a natural step forward, but from \cref{sec:examples} it is clear that the necessary generalizations remain challenging.
Progress in solving Horn clauses~\cite{fedyukovich2019quantified,unno2017automating,hojjat2018eldarica,champion2018hoice,gurfinkel2019science,grebenshchikov2012synthesizing,deangelis2014verifying}
will tie into such work.

Induction for the verification of loops (and also recursive procedures)
occurs in a variety of forms in practice,
specifically when strong specifications are desired and manual effort is acceptable.
In the context of Separation Logic, \citet{tuerk2010local} demonstrates verification rules with contracts, mechanized in the HOL theorem prover,
we refer back to \cref{sec:hoare} for a comparison.
VeriFast offers loop specifications in terms of pre-/postconditions, with first-class support for applying the inductive hypothesis, which was used to solve a challenge with linked trees in the VerifyThis competition~2012 (\citet{jacobs2015solving}).
Tools that embed program verification into a general purpose theorem prover can make use of explicit induction, too, as shown e.g. with KIV (for the same challenge, \citet{ernst:sttt2015}).
Alternatively, loops can be turned into recursive procedures to aid verification~\cite{myreen2009transforming}.

The support for magic wands in~\citet{schwerhoff2015lightweight}
resembles the encoding of contracts as invariants via~\eqref{eq:lift}.
Specifically, \cite[Sec~7]{schwerhoff2015lightweight}
spells this out as
$\mathit{pre}_\mathit{rest} * (\mathit{post}_\mathit{rest} \mathbin{-\!*} \mathit{post}_\mathit{all})$,
albeit this construction does not shed insight into the state variables involved,
lacking the analogue of $sₙ$ in~\eqref{eq:lift}.

A proof system based on coinduction (i.e. forward reasoning)
that generalizes loop verification to any recurring program locations
is~\cite{chen2020towards}, a similar approach to tackle interleavings in concurrency is~\cite{schellhorn2014rgitl}.
\citet{brotherston2005cyclic} takes a similar route to construct cyclic proofs,
implicitly making use of induction.
All three approaches have in common that the inductive property is constructed on the fly instead of being expressed by a fixed predicate or formula up-front.

Recent work on the verification of unstructured assembly programs~\cite{lundberg2020hoare} notes that a negative loop test~$\lnot t$ cannot be assumed by default
in such a setting, as it is the case with the conditions~\cref{def:sm}.
As discussed, our approach can include this conclusion
as part of the loop postcondition, in the absence of $\BREAK$.

The tools participating in SV-COMP~\cite{beyer2020advances} show a strong bias towards invariants.
As an example, the most successful configuration of CPAchecker~\cite{beyer2011cpachecker},
CPA-seq, relies on k-Induction~\cite{donaldson2011software,beyer2018unifying},
which exploits correctness constraints from inside the loop body,
but disregards the constraints after a particular loop for its verification.
Likewise, Property Directed Reachability~\cite{mcmillan2003interpolation}
reasons forward over loops, but it is more precise wrt.~different stages of a computation;
Z3's fixpoint engine is fundamentally based on this idea~\cite{hoder2011muz,komuravelli2013automatic}.

Ultimate Automizer~\cite{heizmann2009refinement} abstracts traces in terms
of automata, which may contain loops.
Whether there is a more fundamental connection between their approach and loop contracts is not quite clear, and we leave this question for future work.
SeaHorn~\cite{gurfinkel2015seahorn} is a verification platform for Horn clause
based verification of C programs, using the invariant approach.

Backwards program analysis, notably and combinations with forward analysis,
has been used in abstraction refinement~\cite{lin2017squeezing,vizel2013intertwined}.
In contrast to here, the idea is to back-propagate information about counterexamples
instead of correctness conditions to derive invariants via interpolation.
Loop summarization in logics for underapproximation~\cite{devries2011reverse,ohearn2019incorrectness} similarly guarantees reachability of certain states,
the underlying ``backwards variants'' are incomparable to loop summaries.

In Spark/GNATprove, loop invariants can be specified to hold anywhere in the loop body~\cite{hoang2015spark}, and there is an equivalent to~$\old$, referring to the state of the loop entry.
The approach is demonstrated with a verification of the prefixsum algorithm,
a tricky challenge from VerifyThis~2012~\cite{huisman2013verifythis}.
The approach bears resemblance to k-Induction~\cite{donaldson2011software},
because correctness conditions are reflected inside the loop, not in the code that follows as with contracts.

\section{Conclusion and Outlook}
\label{sec:conclusion}

This paper presents a concise and accessible formulation
of loop contracts, which generalizes Hoare's~\cite{hoare1969axiomatic}
proof approach for loops using invariants,
and Hehner's refinement-based approach~\cite{hehner1999refinement}.
The presentation sheds light into fundamental properties
of loop contracts, and we prove several novel results
that are not just theoretical but offer immediate guidelines for tool construction:
We characterize, which part of the proof may be expressed using summaries,
and which part must be covered by invariants (\cref{thm:cc}).
Moreover, the approaches can represent each other,
and we give constructive translations between them (\cref{prop:lift,prop:lower}).

We have exemplified the use of contracts versus invariants
on some standard verification tasks.
Both approaches have their respective advantages and disadvantages,
as summarized at the end of \cref{sec:examples}.

A clear path to future work is to develop algorithms that
synthesize loop summaries from procedure postconditions,
similarly to~\citet{furia2010inferring} for invariants,
but potentially exploiting their close correspondence.

Overall, we hope that in the future, verification based on loop-contracts
finds its way into mainstream tools, and helps leverage their possibilities
for those problems where they are beneficial.

\paragraph{Acknowledgement.}
Many thanks to Toby Murray for valuable feedback.
The presentation in \cref{sec:hoare} is part of
Gregor Alexandru's bachelor thesis \cite{alexandru2019}.
The treatment of $\BREAK$ and $\GOTO$ in with loop contracts has been explored by Johannes Blau.
We thank Rustan Leino for the phone number example.

\bibliographystyle{splncsnat}
\bibliography{references}

\end{document}